\documentclass[english,reprint, superscriptaddress, secnumarabic, nofootinbib, nobibnotes, groupedaddress]{revtex4-1}
\usepackage[T1]{fontenc}
\usepackage[latin9]{inputenc}
\usepackage[active]{srcltx}
\usepackage{xcolor}
\usepackage{array}
\usepackage{float}
\usepackage{textcomp}
\usepackage{multirow}
\usepackage{amsmath}
\usepackage{amsthm}
\usepackage{mathdots}
\usepackage[all]{xy}
\PassOptionsToPackage{normalem}{ulem}
\usepackage{ulem}

\makeatletter

\providecommand{\tabularnewline}{\\}
\providecolor{lyxadded}{rgb}{0,0,1}
\providecolor{lyxdeleted}{rgb}{1,0,0}

\theoremstyle{plain}
\newtheorem{thm}{\protect\theoremname}[section]
  \theoremstyle{definition}
  \newtheorem{defn}[thm]{\protect\definitionname}

\usepackage[all]{xy}
\usepackage{morefloats}
\usepackage{placeins} 
\usepackage{lineno}
\usepackage{flushend}
\usepackage{graphicx}
\usepackage{tikz}

\def\frontmatter@abstractheading{}

\makeatletter
\renewcommand{\p@subsection}{}
\renewcommand{\p@subsubsection}{}
\makeatother

\makeatother

\usepackage{babel}
  \providecommand{\definitionname}{Definition}
\providecommand{\theoremname}{Theorem}

\begin{document}

\title{Advanced Analysis of Quantum Contextuality in a Psychophysical Double-Detection
Experiment}

\author{Víctor H. Cervantes }
\email{E-mail: cervantv@purdue.edu}

\author{Ehtibar N. Dzhafarov}
\email{E-mail: ehtibar@purdue.edu}

\affiliation{Purdue University, USA}
\begin{abstract}
The results of behavioral experiments typically exhibit inconsistent
connectedness, i.e., they violate the condition known as ``no-signaling,''
``no-disturbance,'' or ``marginal selectivity.'' This prevents
one from evaluating these experiments in terms of quantum contextuality
if the latter understood traditionally (as, e.g., in the Kochen-Specker
theorem or Bell-type inequalities). The Contextuality-by-Default (CbD)
theory separates contextuality from inconsistent connectedness. When
applied to quantum physical experiments that exhibit inconsistent
connectedness (due to context-dependent errors and/or signaling),
the CbD computations reveal quantum contextuality in spite of this.
When applied to a large body of published behavioral experiments,
the CbD computations reveal no quantum contextuality: all context-dependence
in these experiments is described by inconsistent connectedness alone.
Until recently, however, experimental analysis of contextuality was
confined to so-called cyclic systems of binary random variables. Here,
we present the results of a psychophysical double-detection experiment
that do not form a cyclic system: their analysis requires that we
use a recent modification of CbD, one that makes the class of noncontextual
systems more restricted. Nevertheless our results once again indicate
that when inconsistent connectedness is taken into account, the system
exhibits no contextuality. 

\end{abstract}
\maketitle
In recent years there were many reports of behavioral experiments
\citep{Accardi.2016.Application,Aerts.2014.Quantuma,Aerts.2015.New,Aerts.2015.Spin,Asano.2014.Violation,Bruza.2015.probabilistic,Sozzo.2015.Conjunction,Wang.2014.Context,Zhang.Inpress.Testing,Cervantes.Inpress.Exploration,Dzhafarov.2015.there,Khrennikov.2015.Quantum-like}
aimed at (or interpretable as aimed at) revealing \emph{contextuality}
of the kind predicted by and experimentally confirmed in quantum physics
\citep{Bell.1964.Einstein,Clauser.1969.Proposed,Fine.1982.Hidden,Klyachko.2008.Simple,Kochen.1967.problem,Kurzynski.2012.Entropic,Lapkiewicz.2011.Experimental,Hensen.2015.Loophole-free}.
All known to us behavioral data, however, violate a certain condition
that makes a direct application of the traditional quantum contextuality
analysis impossible. This condition is variously called ``no-signaling''
or ``no-disturbance'' in quantum physics \citep{Bacciagaluppi.2014.Leggett,Bacciagaluppi.2016.Einstein,Kofler.2013.Conditiona,Kurzynski.2014.Fundamental,Ramanathan.2012.Generalized,Cereceda.2000.Quantum,Popescu.1994.Quantum}
and ``marginal selectivity'' in psychology \citep{Dzhafarov.2003.Selective,Zhang.2015.Noncontextuality,Townsend.1989.trichotomy}.
It is a required condition for the traditional quantum contextuality
analysis, even though it is often violated in quantum mechanical experiments
as well (this issue was first systematically discussed in \citep{Adenier.2007.Fair};
see also (\citep{Lapkiewicz.2011.Experimental,Lapkiewicz.2013.Comment,Adenier.2016.Testa}).
The Contextuality-by-Default (CbD) theory \citep{deBarros.2015.Measuring,Dzhafarov.2014.Contextualitya,Dzhafarov.2014.Embedding,Dzhafarov.2015.Contextualitya,Dzhafarov.2015.Random,Dzhafarov.2016.Context-contentb,Dzhafarov.2016.Contextuality-by-Default:,Dzhafarov.2016.Conversations,Dzhafarov.2016.Probabilisticb,Dzhafarov.Inpress.Contextuality-by-Default,Dzhafarov.Inpress.Stochastic}
overcomes this difficulty by proposing a principled way of separating
contextuality proper from \emph{inconsistent connectedness} (the CbD
term for violations of the ``no-signaling'' or ``marginal selectivity''
condition). This theory was used to reanalyze the behavioral experiments
aimed at contextuality, with the conclusion that they provide no evidence
for contextuality \citep{Dzhafarov.2015.there,Dzhafarov.2016.contextuality,Cervantes.Inpress.Exploration,Zhang.Inpress.Testing}:
inconsistent connectedness is the only form of context-dependence
that we have in them. By contrast, when CbD is used to reanalyze a
quantum-mechanical experiment that exhibits inconsistent connectedness
\citep{Lapkiewicz.2011.Experimental}, contextuality proper (on top
of inconsistent connectedness) is established beyond doubt \citep{Kujala.2015.Necessarya}. 

Virtually all experiments aimed at revealing contextuality, both in
quantum physics and in behavioral sciences, deal with a special kind
of systems of random variables, called \emph{cyclic systems} in CbD
\citep{Kujala.2015.Necessarya}. In these systems each property is
measured in precisely two different contexts, and each context contains
two properties being measured together. If, in addition, all random
variables in the system are binary (each indicating presence or absence
of a certain property), then the system is amenable to complete and
exhaustive contextuality analysis \citep{Dzhafarov.2015.Contextualitya,Kujala.2015.Necessarya,Dzhafarov.2016.Context-contentb,Dzhafarov.2016.Contextuality-by-Default:}.
In spite of their prominence in quantum theory, however, it is highly
desirable to extend contextuality analysis beyond the class of cyclic
systems. Many researchers (although not the present authors) find
the lack of contextuality in behavioral data to be a disappointing
negative result. What if this result is due to the fact that cyclic
systems in human behavior are too simple? What if it is ``too easy''
for a cyclic system to be noncontextual? These are valid questions,
and they will have no definite answers until we have a predictive
theory of (at least certain types of) human behavior on a par with
quantum mechanics. 

In the absence of a predictive theory, the only, admittedly imperfect
way of dealing with these considerations is to expand the experimentation
and contextuality analysis to progressively broader classes of systems.
In this paper we make a first step in this direction by analyzing
a psychophysical experiment whose results form a non-cyclic system
of random variables. This experiment was reported previously \citep{Cervantes.Inpress.Exploration},
but its analysis was confined to extracting from it a large number
of cyclic subsystems and showing all of them to be noncontextual.
It is mathematically possible, however, that a system is contextual
with all its cyclic subsystems being noncontextual. 

A satisfactory way to expand the contextuality analysis beyond cyclic
systems was proposed in a recent modification of CbD, dubbed ``CbD
2.0'' \citep{Dzhafarov.2016.Probabilisticb,Dzhafarov.Inpress.Contextuality-by-Default}:
it is essentially the original CbD in which the measurements of the
same property (say, responses to the same stimulus) are analyzed in
pairs only. This modification has compelling reasons behind it, The
main one is that in the modified theory a subsystem of a noncontextual
system is always noncontextual. Another reason is that contextuality
analysis is reduced to the problem of compatibility of two \emph{uniquely
defined} sets of distributions: the empirically known distributions
of context-sharing random variables and the distributions of the ``multimaximal
couplings'' of the random variables measuring the same property in
different contexts. All of this is clarified below (Section \ref{sec:arbitrary}).
The modification in question does not affect the theory of cyclic
systems, so the results mentioned earlier remain unchanged. However,
when it comes to non-cyclic systems, the modification makes the requirements
that a system should satisfy to be noncontextual more stringent. 

The plan of the paper is as follows. In the next two sections we present
the basics of the CbD theory, in the ``CbD 2.0'' version. The discussion
is primarily confined to systems of binary random variables (dichotomic
measurements), both for simplicity and because the double-detection
experiment to be analyzed involves only dichotomic judgments. In Section
\ref{sec:behavior} we apply this theory to the results of our double-detection
experiment. Our conclusion is that in spite of the notion of noncontextuality
we use being more restrictive than in the original version of the
CbD theory, the double detection experiment does not exhibit any contextuality.

\section{Introduction to Contextuality}

\label{sec:intro}

Every experiment results in a system of random variables. In most
physics experiments these random variables are interpreted as measurements
of properties, in most behavioral experiments they are interpreted
as responses to stimuli, such as questions. For brevity we will use
the term ``measurement'' in both meanings (because responding to
a stimulus can always be viewed as a form of measurement). \emph{What}
is being measured therefore is part of the identity of a random variable
representing a measurement. It is referred to as the \emph{content}
of the random variable. The content, however, does not specify a random
variable uniquely, because one and the same content can be measured
under different conditions, referred to as \emph{contexts}. For instance,
if a content $q$ is measured simultaneously with measurements of
other contents, in some cases $q'$ and in other cases $q''$, then
in the former cases the context is $c=\left(q,q'\right)$ and in the
latter ones it is $c'=\left(q,q''\right)$. As in Refs. \citep{Dzhafarov.2016.Context-contentb,Dzhafarov.Inpress.Contextuality-by-Default},
we will write ``conteXt'' and ``conteNt'' to prevent their confusion
in reading. The conteXt and conteNt of a random variable uniquely
identify it within a given system of random variables. So each random
variable in a system is double-indexed, $R_{q}^{c}$. 

According to the CbD theory's main principle \citep{Dzhafarov.2016.Contextuality-by-Default:,Dzhafarov.2014.Contextualitya,Dzhafarov.2016.Context-contentb,Dzhafarov.2016.Conversations,Dzhafarov.2016.Probabilisticb,Dzhafarov.Inpress.Stochastic},
two random variables $R_{q}^{c}$ and $R_{q'}^{c'}$ are jointly distributed
if and only if $c=c'$, i.e., if and only if they are recorded in
the same conteXt. Otherwise they are \emph{stochastically unrelated},
i.e., joint probabilities for them are undefined. This means, in particular,
that any two $R_{q}^{c}$ and $R_{q}^{c'}$ with the same conteNt
in different conteXts are stochastically unrelated (which implies,
among other things, that they can never be considered to be one and
the same random variable). Their individual distributions may be the
same but they need not be. If these distributions are different, the
system exhibits a form of context-dependence. However, in CbD, this
context-dependence by itself does not say that the system is contextual
in the sense related to how this term is used in quantum mechanics.
Rather the difference in the distributions is treated as manifestation
of information/energy flowing to the measurements of conteNt $q$
from elements of the contexts $c,c'$ other than $q$. We will refer
to this transfer of information/energy as \emph{direct cross-influences}.
Thus, if $c=\left(q,q'\right)$ and $c'=\left(q,q''\right)$, the
conteNt $q$ does, of course, directly influence its measurement,
but, with $q$ fixed, the second conteNt in the pair can also affect
this measurement. This can sometimes be attributed to some physical
action of $q'$ or $q''$ upon the process measuring $q$, or (as
another form of information transfer) it can be a form of contextual
bias, a change in the procedure by which $q$ is measured depending
on what else is being measured. 
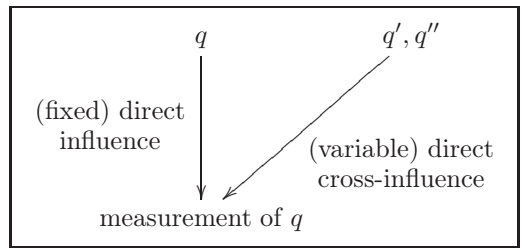
\[
\boxed{\xymatrix{q\ar[dd]_{\begin{array}{c}
\textnormal{(fixed) direct}\\
\textnormal{influence}
\end{array}} & q',q''\ar[ddl]^{\begin{array}{c}
\textnormal{(variable) direct}\\
\textnormal{cross\textnormal{-}influence}
\end{array}}\\
\\
\textnormal{measurement of }q
}
}
\]

The difference between the distributions of $R_{q}^{c}$ and $R_{q}^{c'}$
(equivalently, the strength of the direct cross-influences responsible
for this difference) is measured in CbD by the probability with which
$R_{q}^{c}$ and $R_{q}^{c'}$ could be made to coincide \emph{if
they were jointly distributed}. This means that we consider all \emph{couplings}
of $R_{q}^{c},R_{q}^{c'}$, i.e., the jointly distributed pairs of
random variables $T_{q}^{c},T_{q}^{c'}$ whose respective individual
distributions are the same as those of $R_{q}^{c},R_{q}^{c'}$, and
among these pairs we find the one(s) with the maximal possible probability
of $T_{q}^{c}=T_{q}^{c'}$. The larger this maximal probability, the
closer the two distributions to each other, and the weaker the direct
cross-influences by conteXts $c,c'$ upon the measurement of $q$.
This maximal probability is 1 if and only if the two distributions
are identical, and it is 0 if an only if the two distributions have
disjoint supports. 

Consider now an experiment represented by a system of random variables
$R_{q}^{c}$ with varying $c$ and $q$, and suppose that we have
computed the maximal probability just described for each pair of random
variables that share a conteNt. And we know (or can empirically estimate)
the joint distributions of all random variables that share a conteXt.
Intuitively, quantum contextuality is about whether these computed
maximal probabilities and these empirically defined joint distributions
are mutually compatible. If they are not, then one can say that conteXts
force the random variables sharing conteNts to be more dissimilar
than they are made by direct cross-influences alone. The system then
can be considered \emph{contextual}. 

To understand this without conceptual and technical complications,
consider first a cyclic system of binary random variables \citep{Dzhafarov.2015.Contextualitya,Kujala.2015.Proofa,Kujala.2015.Necessarya}.
It is depicted in Fig. \ref{fig: cyclic n}. The conteXts and conteNts
are such that, with appropriate enumeration, in conteXt $c_{i}$ one
measures precisely two cyclically-successive conteNts $q_{i},q_{i\oplus1}$
(where $i=1,\ldots,n$; $i\oplus1=i+1$ for $i<n$; and $n\oplus1=1$):
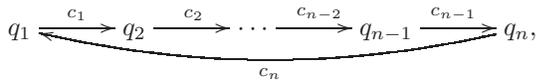
\[
\xymatrix@C=1cm{q_{1}\ar[r]^{c_{1}} & q_{2}\ar[r]^{c_{2}} & \cdots\ar[r]^{c_{n-2}} & q_{n-1}\ar[r]^{c_{n-1}} & q_{n}\ar@/^{1pc}/[llll]^{c_{n}},}
\]
 Each pair $R_{i}^{i},R_{i\oplus1}^{i}$ ($i=1,\ldots,n$) of random
variables sharing a conteXt (within a row in Fig. \ref{fig: cyclic n})
are jointly distributed. Since all the measurements in the system
are binary ($\pm1$), the joint distribution of $R_{j}^{i}$ is uniquely
determined by three probabilities, 
\begin{equation}
\begin{array}{c}
p_{i}^{i}=\Pr\left[R_{i}^{i}=1\right],\;p_{i\oplus1}^{i}=\Pr\left[R_{i\oplus1}^{i}=1\right],\\
\\
p^{i}=\Pr\left[R_{i}^{i}=R_{i\oplus1}^{i}=1\right].
\end{array}\label{eq: bunch cyclic}
\end{equation}
Random variables $R_{i}^{i\ominus1},R_{i}^{i}$ within a column share
a conteNt, and we compute for each such a pair the magnitude of direct
cross-influences, $\max\Pr\left[T_{i}^{i}=T_{i}^{i\ominus1}\right]$,
across all couplings $\left(T_{i}^{i\ominus1},T_{i}^{i}\right)$ of
$R_{i}^{i\ominus1},R_{i}^{i}$: in this case the couplings are the
pairs $\left(T_{i}^{i\ominus1},T_{i}^{i}\right)$ with all possible
values of $\Pr\left[T_{i}^{i}=T_{i}^{i\ominus1}=1\right]$ and with
\begin{equation}
\Pr\left[T_{i}^{i}=1\right]=p_{i}^{i},\;\Pr\left[T_{i}^{i\ominus1}=1\right]=p_{i}^{i\ominus1}.\label{eq: coupling for connection cyclic}
\end{equation}
Here, $i=1,\ldots,n$; $i\ominus1=i-1$ for $i>1$; and $1\ominus1=n$.
The coupling $\left(T_{i}^{i\ominus1},T_{i}^{i}\right)$ with this
property is called \emph{maximal coupling}. It is easy to show \citep{Thorisson.2000.Coupling}
that this maximal coupling always exists and is defined by complementing
(\ref{eq: coupling for connection cyclic}) with
\begin{equation}
p_{i}=\Pr\left[T_{i}^{i}=T_{i}^{i\ominus1}=1\right]=\min\left\{ p_{i}^{i},p_{i}^{i\ominus1}\right\} .\label{eq: connection cyclic}
\end{equation}

The probabilities (\ref{eq: bunch cyclic}) and (\ref{eq: connection cyclic})
are shown in Fig. \ref{fig: cyclic filled}. Note that (\ref{eq: coupling for connection cyclic})
and (\ref{eq: connection cyclic}) uniquely define the joint distribution
of the two random variables $T_{i}^{i\ominus1},T_{i}^{i}$ within
each column of the matrix, in the same way as (\ref{eq: bunch cyclic})
uniquely define the joint distribution of $R_{i}^{i},R_{i\oplus1}^{i}$
within each row of the matrix. The only difference is that the row-wise
joint distributions are empirical reality, whereas the column-wise
joint distributions are constructed artificially to depict the direct
cross-influences. Contextuality in CbD is all about the compatibility
of these column-wise and row-wise joint distributions: the system
is considered noncontextual if all these probabilities can be achieved
within a jointly distributed set of $2n$ random variables. In other
word, we seek a set of jointly distributed random variables $S_{j}^{i}$
replacing the star symbols in Fig. \ref{fig: cyclic n}, such that
\begin{equation}
\begin{array}{ccl}
(i) &  & \Pr\left[S_{i}^{i}=1\right]=p_{i}^{i},\;\Pr\left[S_{i\oplus1}^{i}=1\right]=p_{i\oplus1}^{i},\\
\\
(ii) &  & \Pr\left[S_{i}^{i}=S_{i\oplus1}^{i}=1\right]=p^{i},\\
\\
(iii) &  & \Pr\left[S_{i}^{i}=S_{i}^{i\ominus1}=1\right]=p_{i}=\min\left\{ p_{i}^{i},p_{i}^{i\ominus1}\right\} 
\end{array}\label{eq: cyclic overall}
\end{equation}
The equations (i) and (ii) in (\ref{eq: cyclic overall}) tell us
that the set of the $S_{j}^{i}$-variables we seek is a coupling of
the original random variables $R_{j}^{i}$ arranged row-wise in Fig.
\ref{fig: cyclic n}: in each row the variables $R_{j}^{i}$ have
a well-defined joint distribution, but different rows are stochastically
unrelated, so the coupling ``saws them together'' in a single joint
distribution. The equations (i) and (iii) in (\ref{eq: cyclic overall})
tell us that the set of the $S_{j}^{i}$-variables is a coupling for
the column-wise maximal couplings $T_{j}^{i}$: in each of the columns
the variables $T_{j}^{i}$ have a well-defined joint distribution,
but different columns are stochastically unrelated because the maximal
couplings were computed for each column separately; so the coupling
``saws the columns together'' in a single joint distribution. It
is easy to see that each of these two couplings (of the rows and of
the columns) exists, because the random variables in the different
rows do not overlap, and the same is true for different columns. In
a typical case, each of the two couplings can be constructed in an
infinity of ways, and the question is whether a jointly distributed
set of $2n$ random variables can be simultaneously a coupling for
the rows and for the columns. If the answer to this question is negative,
the conteXts intervene beyond the effect of the direct cross-influences.

\begin{figure}
\begin{centering}
\begin{tabular}{|c|c|c|c|c|c|c|c}
\cline{1-7} 
$\star$ & $\star$ & $ $ & $ $ & $\cdots$ & $ $ & $ $ & $c_{1}$\tabularnewline
\cline{1-7} 
$ $ & $\star$ & $\star$ & $ $ & $\cdots$ & $ $ & $ $ & $c_{2}$\tabularnewline
\cline{1-7} 
$ $ & $ $ & $\star$ & $\star$ & $\cdots$ & $ $ & $ $ & $c_{3}$\tabularnewline
\cline{1-7} 
$\vdots$ & $\vdots$ & $\vdots$ & $\vdots$ & $\iddots$ & $\vdots$ & $\vdots$ & $\vdots$\tabularnewline
\cline{1-7} 
$ $ & $ $ & $ $ & $ $ & $\cdots$ & $\star$ & $\star$ & $c_{n-1}$\tabularnewline
\cline{1-7} 
$\star$ & $ $ & $ $ & $ $ & $\cdots$ & $ $ & $\star$ & $c_{n}$\tabularnewline
\cline{1-7} 
\multicolumn{1}{c}{$q_{1}$} & \multicolumn{1}{c}{$q_{2}$} & \multicolumn{1}{c}{$q_{3}$} & \multicolumn{1}{c}{$q_{4}$} & \multicolumn{1}{c}{$\cdots$} & \multicolumn{1}{c}{$q_{n-1}$} & \multicolumn{1}{c}{$q_{n}$} & $\boxed{\boxed{\mathsf{CYC}}}$\tabularnewline
\end{tabular}
\par\end{centering}
\caption{\label{fig: cyclic n}A cyclic system (shown here for a sufficiently
large $n$, although $n$ can be as small as 2 or 3). The system involves
$n$ conteNts $q_{1},\ldots,q_{n}$ and $n$ conteXts $c_{1},\ldots,c_{n}$.
The star symbol in the $\left(c_{i},q_{j}\right)$-cell indicates
that conteNt $q_{j}$ was measured in conteXt $c_{i}$, and the result
of the measurement is random variable $R_{j}^{i}$; otherwise $q_{j}$
was not measured in $c_{i}$ and the cell is left empty. All $R_{j}^{i}$
are binary random variables, with possible values denoted $+1$ and
$-1$. }
\end{figure}

\begin{figure}
\begin{centering}
\begin{tabular}{|c|c|c|c|c|c|c|c}
\cline{1-7} 
$p_{1}^{1}$ & $p_{2}^{1}$ & $ $ & $ $ & $\cdots$ & $ $ & $ $ & $p^{1}$\tabularnewline
\cline{1-7} 
$ $ & $p_{2}^{2}$ & $p_{3}^{2}$ & $ $ & $\cdots$ & $ $ & $ $ & $p^{2}$\tabularnewline
\cline{1-7} 
$ $ & $ $ & $p_{3}^{3}$ & $p_{4}^{3}$ & $\cdots$ & $ $ & $ $ & $p^{3}$\tabularnewline
\cline{1-7} 
$\vdots$ & $\vdots$ & $\vdots$ & $\vdots$ & $\iddots$ & $\vdots$ & $\vdots$ & $\vdots$\tabularnewline
\cline{1-7} 
$ $ & $ $ & $ $ & $ $ & $\cdots$ & $p_{n-1}^{n-1}$ & $p_{n}^{n-1}$ & $p^{n-1}$\tabularnewline
\cline{1-7} 
$p_{1}^{n-1}$ & $ $ & $ $ & $ $ & $\cdots$ & $ $ & $p_{n}^{n}$ & $p^{n}$\tabularnewline
\cline{1-7} 
\multicolumn{1}{c}{$p_{1}$} & \multicolumn{1}{c}{$p_{2}$} & \multicolumn{1}{c}{$p_{3}$} & \multicolumn{1}{c}{$p$} & \multicolumn{1}{c}{$\cdots$} & \multicolumn{1}{c}{$p_{n-1}$} & \multicolumn{1}{c}{$p_{n}$} & $\boxed{\boxed{\mathsf{CYC}}}$\tabularnewline
\end{tabular}
\par\end{centering}
\caption{\label{fig: cyclic filled}The probability values that characterize
the cyclic system in Fig. \ref{fig: cyclic n} in accordance with
(\ref{eq: bunch cyclic}) and (\ref{eq: connection cyclic}). The
system is noncontextual if there is a set of $2n$ jointly distributed
random variables $\left(S_{j}^{i}:i=1,\ldots,n;j=i\textnormal{ or }j=i\oplus1\right)$
with $\Pr\left[S_{j}^{i}=1\right]=p_{j}^{i}$, $\Pr\left[S_{i}^{i}=S_{i\oplus1}^{i}=1\right]=p^{i}$,
and $\Pr\left[S_{i}^{i}=S_{i}^{i\ominus1}=1\right]=p_{i}=\min\left\{ p_{i}^{i},p_{i}^{i\ominus1}\right\} $. }
\end{figure}

\section{Contextuality in Arbitrary Systems of Binary Measurements}

\label{sec:arbitrary}

Let us discuss now how the analysis just presented extends beyond
cyclic systems. We will continue to assume that all the random variables
in play are binary. 

Consider Fig. \ref{fig: cyclic and X}. The system $\mathsf{X}$ is
not cyclic, as it has three random variables in the first row (conteXt
$c_{1}$) and three random variables in the fourth column (conteNt
$q_{4}$). The number and arrangement of the random variables in a
row, however, is immaterial for the logic of the contextuality analysis.
The joint distribution of $R_{1}^{1},R_{2}^{1},R_{4}^{1}$ in the
first row of $\mathsf{X}$ is uniquely defined empirically. It simply
requires more probabilities than in (\ref{eq: bunch cyclic}) to be
described:
\begin{equation}
\begin{array}{c}
p_{1}^{1}=\Pr\left[R_{1}^{1}=1\right],\;p_{2}^{1}=\Pr\left[R_{2}^{1}=1\right],\\
p_{4}^{1}=\Pr\left[R_{4}^{1}=1\right],\\
\\
p_{12}^{1}=\Pr\left[R_{1}^{1}=R_{2}^{1}=1\right],\\
p_{24}^{1}=\Pr\left[R_{2}^{1}=R_{4}^{1}=1\right],\\
p_{14}^{1}=\Pr\left[R_{1}^{1}=R_{4}^{1}=1\right],\\
\\
p_{124}^{1}=\Pr\left[R_{1}^{1}=R_{2}^{1}=R_{4}^{1}=1\right].
\end{array}\label{eq: first row X}
\end{equation}
Nor does anything change in how one treats the pairs of the conteNt-sharing
random variables in the first three columns: one computes the maximal
coupling for each of these columns. One faces choices, however, when
dealing with the three random variables in the fourth column. What
is the right way of generalizing the maximal coupling in this case?
There is a compelling reason \citep{Dzhafarov.2016.Probabilisticb,Dzhafarov.Inpress.Contextuality-by-Default}
to consider the three conteNt-sharing random variables one pair at
a time, and to compute maximal couplings for them separately. This
means finding a jointly distributed triple $\left(T_{4}^{1},T_{4}^{3},T_{4}^{4}\right)$
whose elements are distributional copies of, respectively, $R_{4}^{1},R_{4}^{3},R_{4}^{4}$,
i.e.,
\begin{equation}
\begin{array}{c}
\Pr\left[T_{4}^{1}=1\right]=p_{4}^{1},\;\Pr\left[T_{4}^{3}=1\right]=p_{4}^{3},\\
\Pr\left[T_{4}^{4}=1\right]=p_{4}^{4},
\end{array}
\end{equation}
such that $\left(T_{4}^{1},T_{4}^{3}\right)$ is the maximal coupling
of $R_{4}^{1},R_{4}^{3}$, $\left(T_{4}^{3},T_{4}^{4}\right)$ is
the maximal coupling of $R_{4}^{3},R_{4}^{4}$, and $\left(T_{4}^{1},T_{4}^{4}\right)$
is the maximal coupling of $R_{4}^{1},R_{4}^{4}$. In terms of probability
values,
\begin{equation}
\begin{array}{c}
\Pr\left[T_{4}^{1}=T_{4}^{3}=1\right]=\min\left\{ p_{4}^{1},p_{4}^{3}\right\} ,\\
\Pr\left[T_{4}^{3}=T_{4}^{4}=1\right]=\min\left\{ p_{4}^{3},p_{4}^{4}\right\} ,\\
\Pr\left[T_{4}^{1}=T_{4}^{4}=1\right]=\min\left\{ p_{4}^{1},p_{4}^{4}\right\} .
\end{array}\label{eq: multi example}
\end{equation}
As shown in \citep{Dzhafarov.Inpress.Contextuality-by-Default,Dzhafarov.2016.Probabilisticb},
such a coupling (called \emph{multimaximal} in CbD ) always exists,
and it is unique (as all the random variables here are binary). The
above-mentioned compelling reason for maximizing the couplings pairwise
is that then, if the system is noncontextual, it will remain noncontextual
after one deletes from it one or more random variables. In other words,
any subsystem of a noncontextual system is noncontextual. This would
not be true, for instance, if we only maximized the value of $\Pr\left[T_{1}^{1}=T_{2}^{1}=T_{4}^{1}=1\right]$.
At the same time, the maximization of $\Pr\left[T_{1}^{1}=T_{2}^{1}=T_{4}^{1}=1\right]$
is achieved ``automatically'' if (\ref{eq: multi example}) is satisfied.
Moreover, one of the equalities in (\ref{eq: multi example}) is redundant
as it can be derived from the other two: if, e.g., $p_{4}^{3}\leq p_{4}^{1}\leq p_{44}$,
then the redundant equality in (\ref{eq: multi example}) is the second
one.Generalizing, we have the following theorem.
\begin{thm}[\citep{Dzhafarov.2016.Probabilisticb,Dzhafarov.Inpress.Contextuality-by-Default}]
\label{thm: multimax}Let $R_{q}^{1},\ldots,R_{q}^{k}$, $k>1$,
be binary ($\pm1$) random variables with conteXts enumerated so that
\[
p_{q}^{1}=\Pr\left[R_{q}^{1}=1\right]\leq\ldots\leq\Pr\left[R_{q}^{k}=1\right]=p_{q}^{k}.
\]
Then there is a unique set of jointly distributed $\left(T_{q}^{1},\ldots,T_{q}^{k}\right)$
such that $\left(T_{q}^{i},T_{q}^{i+1}\right)$ is the maximal coupling
of $R_{q}^{i},R_{q}^{i+1}$, for $i=1,\ldots,k-1$. The coupling $\left(T_{q}^{1},\ldots,T_{q}^{k}\right)$
has the following properties. 

(i) For any subset $\left\{ i_{1},\ldots,i_{m}\right\} \subseteq\left(1,\ldots,k\right)$
with $m\leq k$, $\left(T_{q}^{i_{1}},\ldots,T_{q}^{i_{m}}\right)$
is the maximal coupling of $R_{q}^{i_{1}},\ldots,R_{q}^{i_{m}}$,
i.e., $\Pr\left[T_{q}^{i_{1}}=\ldots=T_{q}^{i_{m}}\right]$ has the
maximal possible value among all couplings of $R_{q}^{i_{1}},\ldots,R_{q}^{i_{m}}$.
In particular, for any $i,j\in\left(1,\ldots,k\right)$, $\left(T_{q}^{i},T_{q}^{j}\right)$
is the maximal coupling of $R_{q}^{i},\ldots,R_{q}^{j}$.

(ii) The distribution of $\left(T_{q}^{i_{1}},\ldots,T_{q}^{i_{m}}\right)$
is defined by {\small{}
\begin{equation}
\begin{array}{c}
\Pr\left[T_{q}^{1}=\ldots=T_{q}^{k}=1\right]=p_{1},\\
\\
\Pr\left[T_{q}^{1}=\ldots=T_{q}^{l}=-1\textnormal{ ; }T_{q}^{l+1}=\ldots=T_{q}^{k}=1\right]=p_{l+1}-p_{l},\\
\textnormal{(for }l=1,\ldots,k-1\textnormal{)}\\
\\
\Pr\left[T_{q}^{1}=\ldots=T_{q}^{k}=-1\right]=1-p_{k},\\
\\
\end{array}\label{eq: multimaximal}
\end{equation}
}with all other combinations of values having probability zero. 
\end{thm}
\begin{figure}
\begin{centering}
\begin{tabular}{|c|c|c|c|lll}
\cline{1-4} 
$\star$ & $\star$ & $ $ & $ $ & $c_{1}$ &  & \tabularnewline
\cline{1-4} 
$ $ & $\star$ & $\star$ & $ $ & $c_{2}$ &  & \tabularnewline
\cline{1-4} 
$ $ & $ $ & $\star$ & $\star$ & $c_{3}$ &  & \tabularnewline
\cline{1-4} 
$\star$ & $ $ & $ $ & $\star$ & $c_{4}$ &  & \tabularnewline
\cline{1-4} 
\multicolumn{1}{c}{$q_{1}$} & \multicolumn{1}{c}{$q_{2}$} & \multicolumn{1}{c}{$q_{3}$} & \multicolumn{1}{c}{$q_{4}$} & $\boxed{\boxed{\mathsf{CYC}_{4}}}$ &  & \tabularnewline
\end{tabular}%
\begin{tabular}{|c|c|c|c|c}
\cline{1-4} 
$\star$ & $\star$ & $ $ & $\star$ & $c_{1}$\tabularnewline
\cline{1-4} 
$ $ & $\star$ & $\star$ & $ $ & $c_{2}$\tabularnewline
\cline{1-4} 
$ $ & $ $ & $\star$ & $\star$ & $c_{3}$\tabularnewline
\cline{1-4} 
$\star$ & $ $ & $ $ & $\star$ & $c_{4}$\tabularnewline
\cline{1-4} 
\multicolumn{1}{c}{$q_{1}$} & \multicolumn{1}{c}{$q_{2}$} & \multicolumn{1}{c}{$q_{3}$} & \multicolumn{1}{c}{$q_{4}$} & $\boxed{\boxed{\mathsf{X}}}$\tabularnewline
\end{tabular}
\par\end{centering}
\caption{\label{fig: cyclic and X}A cyclic system with $n=4$ ($\mathsf{CYC}_{4}$)
and a system $\mathsf{X}$ obtained from $\mathsf{CYC}_{4}$ by adding
to it the random variable $R_{4}^{1}$. }
\end{figure}

Now we can formulate the generalization of the definition of contextuality
given in the previous section.
\begin{defn}
\label{def: contextulaity}A system of binary random variables $R_{q}^{c}$
is noncontextual if there exists a jointly distributed set of (correspondingly
labeled) random variables $S_{q}^{c}$ such that (i) for every conteXt
$c$, the joint distribution of all $S_{q}^{c}$ with this value of
$c$ is identical to the joint distribution of the corresponding $R_{q}^{c}$;
and (ii) for every conteNt $q$, the joint distribution of all $S_{q}^{c}$
with this value of $q$ forms the (unique) multimaximal coupling of
the corresponding $R_{q}^{c}$. 
\end{defn}
The notion of contextuality is, once again, about compatibility of
the uniquely determined row-wise and column-wise distributions. The
row distributions are empirically given, the column distributions
are computed as multimaximal couplings, and the question is whether
it is possible to find a single coupling for both the rows and the
columns. Once again, the logic of the approach is that if the coupling
in question does not exist, it means that the conteXts force some
pairs of the random variables measuring the same conteNt to be more
dissimilar than they are made by direct cross-influences alone \textemdash{}
and the system is therefore contextual.

If a system of random variables turns out to be contextual, one can
compute the degree of its contextuality as the smallest possible total
variation of \emph{quasi-couplings} of this system. A quasi-coupling
differs from a coupling in that the probabilities for its values are
replaced with arbitrary real numbers (not necessarily nonnegative)
that sum to 1. The existence of quasi-couplings for any system and
the uniqueness of the minimum total variation are proved in Ref. \citep{Dzhafarov.2016.Context-contentb}.
We need not discuss this otherwise important topic further because
the experimental results reported below reveal no contextuality. 

\section{Double-Detection Experiment}

\label{sec:behavior}

We now apply the theory just described to the results of a double-detection
experiment. We remind the reader that this experiment was previously
described in Ref. \citep{Cervantes.Inpress.Exploration}, but to keep
this paper self-sufficient we recapitulate the procedural details
below. In Ref. \citep{Cervantes.Inpress.Exploration} the system formed
by the data was analyzed by extracting from it a multitude of cyclic
subsystems. In this paper we analyze the system in its entirety. 

The double-detection experiment is one of only two contextuality-aimed
experiments known to us that uses a within-subject design, i.e., with
probabilities estimated from the responses of a single person to multiple
replications of stimuli. (The other such experiment is the psychophysical
matching one described in Refs. \citep{Dzhafarov.2015.there,Zhang.Inpress.Testing}.)
Most experiments use aggregation of responses obtained from many persons.
The double detection paradigm suggested in \citep{Dzhafarov.2012.Quantuma}
and \citep{Dzhafarov.2013.Probability} provides a framework where
both (in)consistent connectedness and contextuality can be studied
in a manner very similar to how they are studied in quantum-mechanical
systems (or could be studied, because consistent connectedness in
quantum physics is often assumed rather than documented). 

\subsection{Method}

\subsubsection{Participants}

\label{sec:part}The participants were three volunteers, graduate
students at Purdue University, two females and one male (the first
author of this paper), aged around $30$, with normal or corrected
to normal vision. The experimental program was regulated by the Purdue
University's IRB protocol $\#1202011876$. The participants are identified
as $P1-P3$ in the text below.

\subsubsection{Equipment}

\label{sec:apparatus} A personal computer was used with an Intel\textsuperscript{{\scriptsize{}\textregistered{}}}
Core{\scriptsize{}\texttrademark{}} processor running Windows XP,
a $24$-in.\ monitor with a resolution of $1920\times1200$ pixels
(px), and a standard US $104$-key keyboard. The participant's head
was steadied in a chin-rest with forehead support at 90 cm distance
from the monitor; at this distance a pixel on the screen subtended
$62$ sec arc.

\subsubsection{Stimuli}

\label{sec:stim} 
\begin{figure}[tbh]
\centering{}\begin{center}
\begin{tikzpicture}
	\draw[line width=.4] (8,6) circle[radius=1.35];
	\draw[line width=.4] (11.2,6) circle[radius=1.35];
	\filldraw[thick] (8,6) circle[radius=.2pt];
	\filldraw[thick] (11.2,6.04) circle[radius=.2pt];
\end{tikzpicture}
\par\end{center} \caption{An example of the stimulus in the double-detection experiment. In
the left circle the dot is in the center, in the right one it is shifted
4 px upwards. The participant's task was to say, for each of the two
circles, whether the dot was in the center (the answer coded 1) or
off-center (the answer coded -1), irrespective of whether it was shifted
up or down.}
\label{fig:stimulus}
\end{figure}
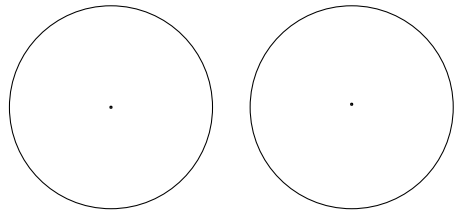

The stimuli presented on the computer screen consisted
of two brightly grey colored circles (RGB 100-100-100) on a black
background, with their centers 320 px apart horizontally, each circle
having the radius of $135$ px and circumference 4 px wide. Each circle
contained a dot of 4 px in diameter in its center or 4 px away from
it, in the upward or downward direction. An example of the stimuli
(in reversed contrast and scaled) is shown in Figure~\ref{fig:stimulus}. 

\subsubsection{Procedure}

\label{sec:proc} In each trial the participant was asked to indicate,
for each circle, whether the dot was in its center or not in the center
(irrespective of in what direction). The responses were given by pressing
in any order and holding together two designated keys, one for each
circle, and the stimuli were displayed until both keys were pressed.
Then, the dots in each circle disappeared, and a ``Press the space
bar to continue'' message appeared above the circles. Pressing the
space bar removed the message, and the next pair of dots appeared
$400$ ms later. (Response times were recorded but not used in the
data analysis.)

Each participant completed nine experimental sessions, each lasting
$30$ minutes and containing about $560$ trials recorded and used
for the analysis, preceded by several practice trials. In each practice
trial the participants received feedback as to whether their response
for each of the two circles was correct or not. No feedback was given
in the non-practice trials. The experimental sessions were preceded
by one to three training sessions, excluded from the analysis.

\subsection{Experimental ConteXts and ConteNts}

\label{sec:expCond} In each of two circles the dot presented could
be located either at its center, or 4 px above the center, or else
4 px under the center. These pairs of locations produce a total of
nine conteXts. During each session, excepting the practice trials,
the dot was presented at the center in a half of the trials, above
the center in a quarter of them, and below the center in the remaining
quarter, for each of the circles. Table~\ref{tab:allocation} presents
the proportions of allocations of trials to each of the 9 conditions.

\begin{figure}
\begin{centering}
\begin{tabular}{|r|r|c|c|c|c|}
\cline{3-5} 
\multicolumn{1}{r}{} &  & \multicolumn{3}{c|}{Right} & \multicolumn{1}{c}{}\tabularnewline
\cline{3-5} 
\multicolumn{1}{r}{} &  & Center $\left(\textnormal{-}c\right)$  & Up $\left(\textnormal{-}u\right)$  & Down $\left(\textnormal{-}d\right)$  & \multicolumn{1}{c}{}\tabularnewline
\hline 
\multirow{3}{*}{Left} & Center $\left(c\textnormal{-}\right)$  & 1/4 $\left(cc\right)$  & 1/8 $\left(cu\right)$  & 1/8 $\left(cd\right)$ & 1/2\tabularnewline
 & Up $\left(u\textnormal{-}\right)$  & 1/8 $\left(uc\right)$  & 1/16 $\left(uu\right)$ & 1/16 $\left(ud\right)$  & 1/4\tabularnewline
 & Down $\left(d\textnormal{-}\right)$  & 1/8 $\left(dc\right)$  & 1/16 $\left(du\right)$ & 1/16 $\left(dd\right)$ & 1/4\tabularnewline
\hline 
\multicolumn{1}{r}{} &  & 1/2 & 1/4 & 1/4 & \multicolumn{1}{c}{}\tabularnewline
\cline{3-5} 
\end{tabular}
\par\end{centering}
\caption{Probabilities with which a trial was allocated to one of the 9 conteXts,
with the notation used for the conteXts and the conteNts: $c$, $u$,
and $d$ denote that the dot is, respectively, in the center, shifted
up, or shifted down. The 9 conteXts are denoted $cc$, $cu$, $du$,
etc., the left (right) symbol indicating the location of the dot in
the left (respectively, right) circle. To denote conteNts, the location
of a dot is shown on the left or on the right with a dash filling
the other side: thus, $c\textnormal{-}$ denotes the dot in the center
of the left circle, $\textnormal{-}d$ denotes the dot shifted down
in the right circle, etc.\label{tab:allocation}}
\end{figure}

For each session, each trial was randomly assigned to one of the conditions
in accordance with Table~\ref{tab:allocation}. The number of experimental
sessions was chosen so that the expected number of (non-practice)
trials in the conditions with lowest probabilities was at least $300$.
This number of observations was chosen based on Refs.~\citep{Cepeda-Cuervo.2008.Intervalos},
whose results show that coverage errors with respect to nominal values
are below $1\%$ for almost all confidence intervals for proportions
with $n>300$.

The system of random variables representing the data is shown in Figure
\ref{fig:positionSystem}.

\begin{figure}[H]
\begin{centering}
{\scriptsize{}}%
\begin{tabular}{c|c|c|c|c|c|c|}
\multicolumn{1}{c}{} & \multicolumn{1}{c}{$c\textnormal{-}$ } & \multicolumn{1}{c}{$\textnormal{-}c$ } & \multicolumn{1}{c}{$u\textnormal{-}$ } & \multicolumn{1}{c}{$\textnormal{-}u$ } & \multicolumn{1}{c}{$d\textnormal{-}$ } & \multicolumn{1}{c}{$\textnormal{-}d$ }\tabularnewline
\cline{2-7} 
$cc$  & $\star$  & $\star$  & $ $  & $ $  & $ $  & $ $ \tabularnewline
\cline{2-7} 
$uc$  & $ $  & $\star$  & $\star$  & $ $  & $ $  & $ $ \tabularnewline
\cline{2-7} 
$uu$  & $ $  & $ $  & $\star$  & $\star$  & $ $  & $ $ \tabularnewline
\cline{2-7} 
$du$  & $ $  & $ $  & $ $  & $\star$  & $\star$  & $ $ \tabularnewline
\cline{2-7} 
$dd$  & $ $  & $ $  & $ $  & $ $  & $\star$  & $\star$ \tabularnewline
\cline{2-7} 
$cu$  & $\star$  & $ $  & $ $  & $\star$  & $ $  & $ $ \tabularnewline
\cline{2-7} 
$ud$  & $ $  & $ $  & $\star$  & $ $  & $ $  & $\star$ \tabularnewline
\cline{2-7} 
$dc$  & $ $  & $\star$  & $ $  & $ $  & $\star$  & $ $ \tabularnewline
\cline{2-7} 
$cd$  & $\star$  & $ $  & $ $  & $ $  & $ $  & $\star$ \tabularnewline
\cline{2-7} 
\end{tabular}
\par\end{centering}{\scriptsize \par}
\caption{The conteNt-conteXt system of measurements for the double detection
experiment. The cell corresponding to context $xy$ and content $z$
(with $z$ being $x\textnormal{-}$ or $\textnormal{-}y$), if it
contains a star, represents the random variable $R_{z}^{xy}$; the
absence of a star means that content $z$ was not measured in context
$xy$. For instance, $xy=cc$ and $z=c\textnormal{-}$ define a random
variable $R_{c\textnormal{-}}^{cc}$. The random variables within
a given row (in the same conteXt) are jointly distributed. In our
design there are two random variables, $R_{x\textnormal{-}}^{xy}$
and $R_{\textnormal{-}y}^{xy}$ in each conteXt $xy$, and their joint
distribution is uniquely defined by three probabilities: $\Pr\left[R_{x\textnormal{-}}^{xy}=1\right]$,
$\Pr\left[R_{\textnormal{-}y}^{xy}=1\right]$, and $\Pr\left[R_{x\textnormal{-}}^{xy}=R_{\textnormal{-}y}^{xy}=1\right]$.
\label{fig:positionSystem}}
\end{figure}

\begin{figure*}
\begin{centering}
{\scriptsize{}}%
\begin{tabular}{|c|c|c|c|c|c|c|cc|}
\hline 
\multicolumn{1}{|c}{{\scriptsize{}$\textnormal{P1}$ }} & \multicolumn{1}{c}{{\scriptsize{}$c\textnormal{-}$ }} & \multicolumn{1}{c}{{\scriptsize{}$\textnormal{-}c$ }} & \multicolumn{1}{c}{{\scriptsize{}$u\textnormal{-}$ }} & \multicolumn{1}{c}{{\scriptsize{}$\textnormal{-}u$ }} & \multicolumn{1}{c}{{\scriptsize{}$d\textnormal{-}$ }} & \multicolumn{1}{c}{{\scriptsize{}$\textnormal{-}d$ }} & {\scriptsize{}$\Pr\left[X=Y=1\right]$ } & {\scriptsize{}$\#\textnormal{ of trials}$ }\tabularnewline
\cline{2-7} 
{\scriptsize{}$cc$ } & {\scriptsize{}$.7175$ } & {\scriptsize{}$.6365$ } & {\scriptsize{}$ $ } & {\scriptsize{}$ $ } & {\scriptsize{}$ $ } & {\scriptsize{}$ $ } & {\scriptsize{}$.5476$ } & {\scriptsize{}$1260$ }\tabularnewline
\cline{2-7} 
{\scriptsize{}$uc$ } & {\scriptsize{}$ $ } & {\scriptsize{}$.5587$ } & {\scriptsize{}$.2476$ } & {\scriptsize{}$ $ } & {\scriptsize{}$ $ } & {\scriptsize{}$ $ } & {\scriptsize{}$.2095$ } & {\scriptsize{}$630$ }\tabularnewline
\cline{2-7} 
{\scriptsize{}$uu$ } & {\scriptsize{}$ $ } & {\scriptsize{}$ $ } & {\scriptsize{}$.5238$ } & {\scriptsize{}$.4857$ } & {\scriptsize{}$ $ } & {\scriptsize{}$ $ } & {\scriptsize{}$.3746$ } & {\scriptsize{}$315$ }\tabularnewline
\cline{2-7} 
{\scriptsize{}$du$ } & {\scriptsize{}$ $ } & {\scriptsize{}$ $ } & {\scriptsize{}$ $ } & {\scriptsize{}$.0444$ } & {\scriptsize{}$.7810$ } & {\scriptsize{}$ $ } & {\scriptsize{}$.0286$ } & {\scriptsize{}$315$ }\tabularnewline
\cline{2-7} 
{\scriptsize{}$dd$ } & {\scriptsize{}$ $ } & {\scriptsize{}$ $ } & {\scriptsize{}$ $ } & {\scriptsize{}$ $ } & {\scriptsize{}$.7556$ } & {\scriptsize{}$.6508$ } & {\scriptsize{}$.5714$ } & {\scriptsize{}$315$ }\tabularnewline
\cline{2-7} 
{\scriptsize{}$cu$ } & {\scriptsize{}$.8095$ } & {\scriptsize{}$ $ } & {\scriptsize{}$ $ } & {\scriptsize{}$.2302$ } & {\scriptsize{}$ $ } & {\scriptsize{}$ $ } & {\scriptsize{}$.2175$ } & {\scriptsize{}$630$ }\tabularnewline
\cline{2-7} 
{\scriptsize{}$ud$ } & {\scriptsize{}$ $ } & {\scriptsize{}$ $ } & {\scriptsize{}$.0762$ } & {\scriptsize{}$ $ } & {\scriptsize{}$ $ } & {\scriptsize{}$.4571$ } & {\scriptsize{}$.0571$ } & {\scriptsize{}$315$ }\tabularnewline
\cline{2-7} 
{\scriptsize{}$dc$ } & {\scriptsize{}$ $ } & {\scriptsize{}$.3032$ } & {\scriptsize{}$ $ } & {\scriptsize{}$ $ } & {\scriptsize{}$.7937$ } & {\scriptsize{}$ $ } & {\scriptsize{}$.2778$ } & {\scriptsize{}$630$ }\tabularnewline
\cline{2-7} 
{\scriptsize{}$cd$ } & {\scriptsize{}$.4063$ } & {\scriptsize{}$ $ } & {\scriptsize{}$ $ } & {\scriptsize{}$ $ } & {\scriptsize{}$ $ } & {\scriptsize{}$.6349$ } & {\scriptsize{}$.3730$ } & {\scriptsize{}$630$ }\tabularnewline
\cline{2-7} 
\multicolumn{1}{|c}{{\scriptsize{}$\Pr\left[A=B=1\right]$ }} & \multicolumn{1}{c}{{\scriptsize{}$.7175$ }} & \multicolumn{1}{c}{{\scriptsize{}$.5587$ }} & \multicolumn{1}{c}{{\scriptsize{}$.2476$ }} & \multicolumn{1}{c}{{\scriptsize{}$.0444$ }} & \multicolumn{1}{c}{{\scriptsize{}$.7556$ }} & \multicolumn{1}{c}{{\scriptsize{}$.4571$ }} &  & \tabularnewline
\multicolumn{1}{|c}{{\scriptsize{}$\Pr\left[B=C=1\right]$ }} & \multicolumn{1}{c}{{\scriptsize{}$.4063$ }} & \multicolumn{1}{c}{{\scriptsize{}$.3032$ }} & \multicolumn{1}{c}{{\scriptsize{}$.0762$ }} & \multicolumn{1}{c}{{\scriptsize{}$.0444$ }} & \multicolumn{1}{c}{{\scriptsize{}$.7556$ }} & \multicolumn{1}{c}{{\scriptsize{}$.4571$ }} &  & \tabularnewline
\multicolumn{1}{|c}{{\scriptsize{}$\Pr\left[A=C=1\right]$ }} & \multicolumn{1}{c}{{\scriptsize{}$.4063$ }} & \multicolumn{1}{c}{{\scriptsize{}$.3032$ }} & \multicolumn{1}{c}{{\scriptsize{}$.0762$ }} & \multicolumn{1}{c}{{\scriptsize{}$.2302$ }} & \multicolumn{1}{c}{{\scriptsize{}$.7810$ }} & \multicolumn{1}{c}{{\scriptsize{}$.6349$ }} &  & \tabularnewline
\hline 
\end{tabular}
\par\end{centering}{\scriptsize \par}
\caption{Empirical data (relative frequencies) for the conteNt-conteXt system
in Fig. \ref{fig:positionSystem} for participant P1. For every conteXt
$xy$ and every conteNt $z$ measured in $xy$ (either $x\textnormal{-}$
or $\textnormal{-}y$), the cell for $R_{z}^{xy}$ contains the frequency
estimate of $\Pr\left[R_{z}^{xy}=1\right]$; the right margins of
the row for $xy$ shows the frequency estimate of $\Pr\left[R_{x\textnormal{-}}^{xy}=R_{\textnormal{-}y}^{xy}=1\right]$
and the total number of measurements in this conteXt. Since $xy$
and $z$ vary, the column for joint probabilities denotes the two
random variables by $X=R_{x\textnormal{-}}^{xy}$ and $Y=R_{\textnormal{-}y}^{xy}$.
The bottom margins in the column for conteNt $z$ show the three frequency
estimates of the maximal values of $\Pr\left[R_{z}^{xy}=R_{z}^{uv}=1\right]$,
$\Pr\left[R_{z}^{uv}=R_{z}^{st}=1\right]$, and $\Pr\left[R_{z}^{xy}=R_{z}^{st}=1\right]$
(where $xy,uv,st$ are three conteXts in which $z$ was measured).
To make notation compact, the three random variables in each column
are labeled $A,B,C$ (from top down), and the three probabilities
are shown as $\Pr\left[A=B=1\right]$, $\Pr\left[B=C=1\right]$, and
$\Pr\left[A=C=1\right]$ (one of which is always redundant but shown
for completeness).\label{fig:data positionSystem-1-1}}
\end{figure*}

\begin{figure*}
\begin{centering}
{\scriptsize{}}%
\begin{tabular}{|c|c|c|c|c|c|c|cc|}
\hline 
\multicolumn{1}{|c}{{\scriptsize{}$\textnormal{P2}$ }} & \multicolumn{1}{c}{{\scriptsize{}$c\textnormal{-}$ }} & \multicolumn{1}{c}{{\scriptsize{}$\textnormal{-}c$ }} & \multicolumn{1}{c}{{\scriptsize{}$u\textnormal{-}$ }} & \multicolumn{1}{c}{{\scriptsize{}$\textnormal{-}u$ }} & \multicolumn{1}{c}{{\scriptsize{}$d\textnormal{-}$ }} & \multicolumn{1}{c}{{\scriptsize{}$\textnormal{-}d$ }} & {\scriptsize{}$\Pr\left[X=Y=1\right]$ } & {\scriptsize{}$\#\textnormal{ of trials}$ }\tabularnewline
\cline{2-7} 
{\scriptsize{}$cc$ } & {\scriptsize{}$.8659$ } & {\scriptsize{}$.7841$ } & {\scriptsize{}$ $ } & {\scriptsize{}$ $ } & {\scriptsize{}$ $ } & {\scriptsize{}$ $ } & {\scriptsize{}$.6746$ } & {\scriptsize{}$1260$ }\tabularnewline
\cline{2-7} 
{\scriptsize{}$uc$ } & {\scriptsize{}$ $ } & {\scriptsize{}$.7619$ } & {\scriptsize{}$.3968$ } & {\scriptsize{}$ $ } & {\scriptsize{}$ $ } & {\scriptsize{}$ $ } & {\scriptsize{}$.1968$ } & {\scriptsize{}$630$ }\tabularnewline
\cline{2-7} 
{\scriptsize{}$uu$ } & {\scriptsize{}$ $ } & {\scriptsize{}$ $ } & {\scriptsize{}$.5556$ } & {\scriptsize{}$.5841$ } & {\scriptsize{}$ $ } & {\scriptsize{}$ $ } & {\scriptsize{}$.3746$ } & {\scriptsize{}$315$ }\tabularnewline
\cline{2-7} 
{\scriptsize{}$du$ } & {\scriptsize{}$ $ } & {\scriptsize{}$ $ } & {\scriptsize{}$ $ } & {\scriptsize{}$.6317$ } & {\scriptsize{}$.1333$ } & {\scriptsize{}$ $ } & {\scriptsize{}$.0254$ } & {\scriptsize{}$315$ }\tabularnewline
\cline{2-7} 
{\scriptsize{}$dd$ } & {\scriptsize{}$ $ } & {\scriptsize{}$ $ } & {\scriptsize{}$ $ } & {\scriptsize{}$ $ } & {\scriptsize{}$.2413$ } & {\scriptsize{}$.2032$ } & {\scriptsize{}$.1175$ } & {\scriptsize{}$315$ }\tabularnewline
\cline{2-7} 
{\scriptsize{}$cu$ } & {\scriptsize{}$.8508$ } & {\scriptsize{}$ $ } & {\scriptsize{}$ $ } & {\scriptsize{}$.4587$ } & {\scriptsize{}$ $ } & {\scriptsize{}$ $ } & {\scriptsize{}$.3444$ } & {\scriptsize{}$630$ }\tabularnewline
\cline{2-7} 
{\scriptsize{}$ud$ } & {\scriptsize{}$ $ } & {\scriptsize{}$ $ } & {\scriptsize{}$.6127$ } & {\scriptsize{}$ $ } & {\scriptsize{}$ $ } & {\scriptsize{}$.1111$ } & {\scriptsize{}$.0063$ } & {\scriptsize{}$315$ }\tabularnewline
\cline{2-7} 
{\scriptsize{}$dc$ } & {\scriptsize{}$ $ } & {\scriptsize{}$.8905$ } & {\scriptsize{}$ $ } & {\scriptsize{}$ $ } & {\scriptsize{}$.1667$ } & {\scriptsize{}$ $ } & {\scriptsize{}$.1476$ } & {\scriptsize{}$630$ }\tabularnewline
\cline{2-7} 
{\scriptsize{}$cd$ } & {\scriptsize{}$.9429$ } & {\scriptsize{}$ $ } & {\scriptsize{}$ $ } & {\scriptsize{}$ $ } & {\scriptsize{}$ $ } & {\scriptsize{}$.0683$ } & {\scriptsize{}$.0571$ } & {\scriptsize{}$630$ }\tabularnewline
\cline{2-7} 
\multicolumn{1}{|c}{{\scriptsize{}$\Pr\left[A=B=1\right]$ }} & \multicolumn{1}{c}{{\scriptsize{}$.8508$ }} & \multicolumn{1}{c}{{\scriptsize{}$.7619$ }} & \multicolumn{1}{c}{{\scriptsize{}$.3968$ }} & \multicolumn{1}{c}{{\scriptsize{}$.5556$ }} & \multicolumn{1}{c}{{\scriptsize{}$.1333$ }} & \multicolumn{1}{c}{{\scriptsize{}$.1111$ }} &  & \tabularnewline
\multicolumn{1}{|c}{{\scriptsize{}$\Pr\left[B=C=1\right]$ }} & \multicolumn{1}{c}{{\scriptsize{}$.8508$ }} & \multicolumn{1}{c}{{\scriptsize{}$.7619$ }} & \multicolumn{1}{c}{{\scriptsize{}$.5556$ }} & \multicolumn{1}{c}{{\scriptsize{}$.4587$ }} & \multicolumn{1}{c}{{\scriptsize{}$.1667$ }} & \multicolumn{1}{c}{{\scriptsize{}$.0683$ }} &  & \tabularnewline
\multicolumn{1}{|c}{{\scriptsize{}$\Pr\left[A=C=1\right]$ }} & \multicolumn{1}{c}{{\scriptsize{}$.8659$ }} & \multicolumn{1}{c}{{\scriptsize{}$.7841$ }} & \multicolumn{1}{c}{{\scriptsize{}$.3968$ }} & \multicolumn{1}{c}{{\scriptsize{}$.4587$ }} & \multicolumn{1}{c}{{\scriptsize{}$.1333$ }} & \multicolumn{1}{c}{{\scriptsize{}$.0683$ }} &  & \tabularnewline
\hline 
\end{tabular}
\par\end{centering}{\scriptsize \par}
\caption{Empirical data (frequencies) for the conteNt-conteXt system in Fig.
\ref{fig:positionSystem} for participant P2. The rest is as in Fig.
\ref{fig:data positionSystem-1-1}.\label{fig:data positionSystem-1-1-1}}
\end{figure*}

\begin{figure*}
\begin{centering}
{\scriptsize{}}%
\begin{tabular}{|c|c|c|c|c|c|c|cc|}
\hline 
\multicolumn{1}{|c}{{\scriptsize{}$\textnormal{P3}$ }} & \multicolumn{1}{c}{{\scriptsize{}$c\textnormal{-}$ }} & \multicolumn{1}{c}{{\scriptsize{}$\textnormal{-}c$ }} & \multicolumn{1}{c}{{\scriptsize{}$u\textnormal{-}$ }} & \multicolumn{1}{c}{{\scriptsize{}$\textnormal{-}u$ }} & \multicolumn{1}{c}{{\scriptsize{}$d\textnormal{-}$ }} & \multicolumn{1}{c}{{\scriptsize{}$\textnormal{-}d$ }} & {\scriptsize{}$\Pr\left[X=Y=1\right]$ } & {\scriptsize{}$\#\textnormal{ of trials}$ }\tabularnewline
\cline{2-7} 
{\scriptsize{}$cc$ } & {\scriptsize{}$.6791$ } & {\scriptsize{}$.5973$ } & {\scriptsize{}$ $ } & {\scriptsize{}$ $ } & {\scriptsize{}$ $ } & {\scriptsize{}$ $ } & {\scriptsize{}$.3654$ } & {\scriptsize{}$1259$ }\tabularnewline
\cline{2-7} 
{\scriptsize{}$uc$ } & {\scriptsize{}$ $ } & {\scriptsize{}$.8302$ } & {\scriptsize{}$.1349$ } & {\scriptsize{}$ $ } & {\scriptsize{}$ $ } & {\scriptsize{}$ $ } & {\scriptsize{}$.0905$ } & {\scriptsize{}$630$ }\tabularnewline
\cline{2-7} 
{\scriptsize{}$uu$ } & {\scriptsize{}$ $ } & {\scriptsize{}$ $ } & {\scriptsize{}$.2548$ } & {\scriptsize{}$.1688$ } & {\scriptsize{}$ $ } & {\scriptsize{}$ $ } & {\scriptsize{}$.0732$ } & {\scriptsize{}$314$ }\tabularnewline
\cline{2-7} 
{\scriptsize{}$du$ } & {\scriptsize{}$ $ } & {\scriptsize{}$ $ } & {\scriptsize{}$ $ } & {\scriptsize{}$.1460$ } & {\scriptsize{}$.3746$ } & {\scriptsize{}$ $ } & {\scriptsize{}$.0127$ } & {\scriptsize{}$315$ }\tabularnewline
\cline{2-7} 
{\scriptsize{}$dd$ } & {\scriptsize{}$ $ } & {\scriptsize{}$ $ } & {\scriptsize{}$ $ } & {\scriptsize{}$ $ } & {\scriptsize{}$.3460$ } & {\scriptsize{}$.4127$ } & {\scriptsize{}$.1397$ } & {\scriptsize{}$315$ }\tabularnewline
\cline{2-7} 
{\scriptsize{}$cu$ } & {\scriptsize{}$.8381$ } & {\scriptsize{}$ $ } & {\scriptsize{}$ $ } & {\scriptsize{}$.0746$ } & {\scriptsize{}$ $ } & {\scriptsize{}$ $ } & {\scriptsize{}$.0524$ } & {\scriptsize{}$630$ }\tabularnewline
\cline{2-7} 
{\scriptsize{}$ud$ } & {\scriptsize{}$ $ } & {\scriptsize{}$ $ } & {\scriptsize{}$.1178$ } & {\scriptsize{}$ $ } & {\scriptsize{}$ $ } & {\scriptsize{}$.3917$ } & {\scriptsize{}$.0159$ } & {\scriptsize{}$314$ }\tabularnewline
\cline{2-7} 
{\scriptsize{}$dc$ } & {\scriptsize{}$ $ } & {\scriptsize{}$.6714$ } & {\scriptsize{}$ $ } & {\scriptsize{}$ $ } & {\scriptsize{}$.2921$ } & {\scriptsize{}$ $ } & {\scriptsize{}$.1127$ } & {\scriptsize{}$630$ }\tabularnewline
\cline{2-7} 
{\scriptsize{}$cd$ } & {\scriptsize{}$.6968$ } & {\scriptsize{}$ $ } & {\scriptsize{}$ $ } & {\scriptsize{}$ $ } & {\scriptsize{}$ $ } & {\scriptsize{}$.3238$ } & {\scriptsize{}$.1746$ } & {\scriptsize{}$630$ }\tabularnewline
\cline{2-7} 
\multicolumn{1}{|c}{{\scriptsize{}$\Pr\left[A=B=1\right]$ }} & \multicolumn{1}{c}{{\scriptsize{}$.6791$ }} & \multicolumn{1}{c}{{\scriptsize{}$.5973$ }} & \multicolumn{1}{c}{{\scriptsize{}$.1349$ }} & \multicolumn{1}{c}{{\scriptsize{}$.1460$ }} & \multicolumn{1}{c}{{\scriptsize{}$.3460$ }} & \multicolumn{1}{c}{{\scriptsize{}$.3917$ }} &  & \tabularnewline
\multicolumn{1}{|c}{{\scriptsize{}$\Pr\left[B=C=1\right]$ }} & \multicolumn{1}{c}{{\scriptsize{}$.6968$ }} & \multicolumn{1}{c}{{\scriptsize{}$.6714$ }} & \multicolumn{1}{c}{{\scriptsize{}$.1178$ }} & \multicolumn{1}{c}{{\scriptsize{}$.0746$ }} & \multicolumn{1}{c}{{\scriptsize{}$.2921$ }} & \multicolumn{1}{c}{{\scriptsize{}$.3238$ }} &  & \tabularnewline
\multicolumn{1}{|c}{{\scriptsize{}$\Pr\left[A=C=1\right]$ }} & \multicolumn{1}{c}{{\scriptsize{}$.6791$ }} & \multicolumn{1}{c}{{\scriptsize{}$.5973$ }} & \multicolumn{1}{c}{{\scriptsize{}$.1178$ }} & \multicolumn{1}{c}{{\scriptsize{}$.0746$ }} & \multicolumn{1}{c}{{\scriptsize{}$.2921$ }} & \multicolumn{1}{c}{{\scriptsize{}$.3238$ }} &  & \tabularnewline
\hline 
\end{tabular}
\par\end{centering}{\scriptsize \par}
\caption{Empirical data (frequencies) for the conteNt-conteXt system in Fig.
\ref{fig:positionSystem} for participant P3. The rest is as in Fig.
\ref{fig:data positionSystem-1-1}.\label{fig:data positionSystem-1-1-2}}
\end{figure*}

\subsection{Results}

\label{sec:results}

The results are shown in Figs. \ref{fig:data positionSystem-1-1},
\ref{fig:data positionSystem-1-1-1}, and \ref{fig:data positionSystem-1-1-2},
one for each of the three participants. Each row, together with its
margins, specifies an empirical estimate of the joint distribution
of the two random variables sharing the corresponding conteXt. This
distribution is shown in the format
\[
\Pr\left[X=1\right],\Pr\left[Y=1\right],\Pr\left[X=Y=1\right],
\]
where $X$ and $Y$ are the two variables in the same row. Each column,
together with its margins, shows an empirical estimate of the multimaximal
coupling of the three random variables sharing the corresponding conteNt.
The distribution of the coupling is shown in the format
\[
\begin{array}{c}
\Pr\left[A=1\right]\\
\Pr\left[B=1\right]\\
\Pr\left[C=1\right]\\
\Pr\left[A=B=1\right]\\
\Pr\left[B=C=1\right]\\
\Pr\left[A=C=1\right]
\end{array},
\]
where $A,B,C$ are the three random variables in the same column listed
from top down. The analysis of contextuality consists in considering
a set of jointly distributed 18 binary random variables (corresponding
to the star symbols in Fig. \ref{fig:positionSystem}), and determining
whether the $2^{18}$ values of this set can be assigned probabilities
that sum to the probabilities whose empirical estimates are shown
in the data matrices (Figs. \ref{fig:data positionSystem-1-1}, \ref{fig:data positionSystem-1-1-1},
and \ref{fig:data positionSystem-1-1-2}). This is a standard linear
programing task,
\[
\underset{46\times2^{18}}{\mathbf{M}}\overset{2^{18}\times1}{\mathbf{Q}}=\underset{46\times1}{\mathbf{P}},\;\mathbf{Q}>0\;\textnormal{(componentwise).}
\]
 The number of the rows in $\mathbf{M}$ and $\mathbf{P}$ (i.e.,
the number of linear constraints imposed on $\mathbf{Q}$) is the
number of the probability estimates shown in each of the data matrices
(45) plus the constraint that ensures that all the $2^{18}$ probabilities
in $\mathbf{Q}$ sum to 1. (The number of the probability estimates
could be reduced from 45 to 39, because one of the three marginal
probabilities for each column could be eliminated. We did not, however,
make use of this small reduction in our computations.) The linear
programing was performed by using the GLPK (GNU Linear Programming
Kit) package (version 4.6; \citep{Makhorin.2012.GLPK}) and the R
interface to the package (Rglpk, version 0.6-1; \citep{Theussl.2015.R/GNU}).

The outcome of the analysis is that, for all three participants, the
system of linear equations has a solution in nonnegative values \textemdash that
is, the data matrices in Figs. \ref{fig:data positionSystem-1-1},
\ref{fig:data positionSystem-1-1-1}, and \ref{fig:data positionSystem-1-1-2}
describe noncontextual systems of random variables. Note that in this
case the empirical estimates were fit by the solution precisely, eliminating
the need for statistical analysis. 

\section{Conclusion}

\label{sec:conclusions}

The experiment presented in this paper illustrates the use of the
double factorial paradigm in the search of contextuality in behavioral
systems, namely in the responses of human observers in a double-detection
task. This paradigm provides the closest analogue in psychophysical
research to the Alice-Bob EPR/Bohm paradigm \citep{Bell.1964.Einstein,Fine.1982.Hidden,Clauser.1969.Proposed}.
We have found that for the participants in the study there was no
evidence of contextuality in their responses. These results add to
the existing evidence that points towards lack of contextuality in
behavioral data \citep{Dzhafarov.2015.there,Zhang.Inpress.Testing,Dzhafarov.2016.contextuality,Cervantes.Inpress.Exploration}.
The present result is in fact stronger than the previous ones, as
it uses a more stringent than before criterion of noncontextuality.
This criterion is based on multimaximality rather than on the simple
maximality of the couplings in cyclic systems. However, we should
emphasize that in the absence of a predictive theory on a par with
quantum mechanics, no failure to find contextuality in even a large
number of experiments can be safely generalized: contextuality may
very well be found under as yet unexplored modifications of experimental
conditions. Consider, e.g., the Alice-Bob EPR/Bohm paradigm, and imagine
that we have no theory that could guide us in choosing the specific
axes along which Alice and Bob are to measure the spins in their respective
particles. It would be rather unlikely to hit at a ``right'' combination
of the angles by pure chance, and after numerous failures one could
very well conclude, in this case wrongly, that contextuality is absent
in this paradigm. More work is needed.

\paragraph*{Acknowledgments.}

This research has been supported by AFOSR grant FA9550-14-1-0318.

\bibliographystyle{apa}
\bibliography{doubleDetectionPaper}

\end{document}